\documentstyle[12pt,epsf,epsfig,wrapfig,lscape,amsmath,amssymb,
amsfonts,amsbsy,graphics,lscape,mathrsfs,dsfont,color,indentfirst]{article}
\newcommand{\vs}{\mbox{$\vphantom{{\bigl|^{d}}^d}$}}

\begin{document}

\begin{center}
{\bfseries PHYSICAL PROGRAMM AND ACCELERATION OF POLARIZED LIGHT
NUCLEI BEAMS  AT JINR NUCLOTRON}

\vskip 5mm S. Vokal$^{1}$,A.D. Kovalenko$^{1}$,A.M.
Kondratenko$^{2}$,M.A. Kondratenko$^{2}$,V.A. Mikhailov$^{1}$,
\underline{Yu.N. Filatov}$^{1}$ and \underline{S.S.
Shimanskiy}$^{1 \dag}$

\vskip 5mm
{\small
(1) {\it
JINR, Dubna, Russia
},
(2) {\it
TPO "Zaryad", Novosibirsk, Russia
}\\
$\dag$ {\it
E-mail: shimanskiy@jinr.ru
}}
\end{center}

\vskip 5mm
\begin{abstract}
The physical spin program at high $p_T$ region and energies
$s^{1/2}_{NN} \sim 10~GeV$ is discussed. It's shown that
cumulative processes, color transparency problem and polarization
phenomenons directly connect with properties new form of the
nuclear matter as Color Quark Condensate(CQC). Studies of CQC one
of the most important physical problem and can be realized using
polarized ion beams at JINR nuclotron-M (and in future at NICA).
The calculations of  spin resonance strengthes in the linear
approximation for p, d, t and $^3He$ beams in the JINR nuclotron
are presented. The methods to preserve the degree of polarization
during crossing the spin resonances are examined. The method of
matching the direction of polarization vector during the beam
injection in to the ring of the nuclotron is given. These methods
of spin resonance crossing can be used to accelerate polarized
beams in the other cyclic accelerators.

\end{abstract}

\vskip 8mm In the recent decades occurred the radical revision
our understanding of  forms of the nuclear matter which can be
realized at different temperatures and densities \cite{McLerran}.
Nowadays it is predicted that at low temperatures and high
densities the nuclear matter is formed completely the new form,
in which the dominant role play the constituent quarks. This
state can be named as a Quark Color Condensate (QCC). The
properties of this form of nuclear matter determine the physical
properties of matter in the center of massive stars and,
possibly, it is directly connected with the riddles of the
explosions of supernovas. The discovered enormous magnetic fields
in stars (up to $\sim~10^{17}$~T) can lead to the fact that the
Quark Color Condensate (QCC) will be polarized. Therefore the
polarization characteristics of super-dense nuclear matter not
only are interesting by themselves, but they have important
significance to developing the theory of evolution of massive
stars.

Is it possible to obtain nuclear matter at the high densities and
low temperatures in a laboratory? Studies of cumulative
(subthreshold) processes have shown that we observe the
processes, in which nuclear matter exist at low temperatures and
densities which exceed the ordinary nuclear (hadron) density up
to ten times \cite{Shimanskiy}. The density three times greater
then ordinary density  was observed in the processes of the deep
inelastic scattering (DIS) of electrons on the nuclei in JLAB
\cite{Egiyan}. Studies of cumulative processes and DIS processes
have shown that the high density state with a certain probability
exists in the ordinary nuclei (the fact that it is not the product
of compression during the collision was shown by study of the
special features of cumulative processes and lepton DIS processes
at {\it x} up to 3, because lepton cannot compress the nuclear
matter).

It means that in the nuclear matter exist nucleon clusters
(Blokhintsev had named its as fluctons) with the density several
times higher than usual and there is no energy gap for the
transition to the QCC phase. Most likely in the region at low
temperature and high density for the nuclear matter there is not
first-order transition. If we take additionally in to account the
absence of the first-order transition in the region of high
temperatures it can be considered as the indication for the
nuclear matter generally there are not regions of the first-order
transition. This is a picture of the phase transition of nuclear
matter which was popular in the 90's years of the last century
\cite{McLerran}.

The high $p_T$ processes (region of $x_T\sim 1$) deal with the
high density of the nuclear (hadron) matter too.  The color
transparency (CT) (observed for the first time in 1988) \cite{CT}
and elastic $p-p$ cross-sections in the singlet and triplet spin
states at angles $90^o_{cm}$ (middle of 70th) \cite{Krisch} may
be the most interesting phenomenons.

The cumulative processes and processes with high $p_T $ in the
range of energies up to $ \sqrt {s _ {NN}} \sim 10~GeV $ is
possible to describe well using phenomenological approaches based
on the constituent picture only not polarization characteristics.
However till now we can say that there are not complete
understanding ("microscopic" models) of the nature of discovered
effects. Especially difficultly to explain the nature of
polarization effects. It means that there are very poor
understanding of properties of the nuclear matter at high
densities and low temperatures. Very important properties all
these phenomenons that its not vanish at high energy region.
Moreover some features very close to new phenomenons.  Let us
compare CT data \cite{CT} with data from RHIC for the so-called
"jet quenching" effect (Figure 1). We can see very close shape of the CT
data and the RHIC data. That's why we can say that the nature of
high $p_T$ suppression at RHIC directly connect with the nature of
CT phenomena.

Before we have said that the cumulative effects and high $p_T$
effects have been discovered in the energy range up to
$\sqrt {s_{NN}}\sim 10~GeV$. JINR nuclotron is the accelerator of
relativistic nuclei which works and continues to be improved in
the V.I. Veksler and A.M. Baldin Laboratory of  high
energies(LHE). The accelerator uses the magnets with
superconductor coils developed in LHE and has been created to
work with proton beams up to energy $12~GeV$ and  nuclei up to
$6~AGeV$. In JINR is discussing plan to built new collider NICA
with maximal energy $\sqrt{s_{NN}}=9~GeV$. The first stage
to NICA project will be upgrade of the nuclotron to the
nuclotron-M. Polarized light ion beams will be important part of
this new project. With polarized ion beams we will have real
possibility to resolve many problems connected with CQC
properties there are:

\begin{itemize}
\renewcommand{\labelitemi}{--}
\setlength{\itemsep}{0mm}
\item
resolve the "spin crisis" of 70's using complete set polarized states \\
($p\uparrow-p\uparrow,p\uparrow-n\uparrow,n\uparrow-n\uparrow$,...);
\item
understand the nature of color transparency phenomenon \\
($p\uparrow-A, p\uparrow-^3\!He(d)\uparrow$);
\item
understand the nature of cumulative(subthreshold) particle production;
\item
the first time study the properties of polarized nuclear matter \\
($d\uparrow-d\uparrow,^3\!He\uparrow-^3\!He~\uparrow$).

\end{itemize}

\begin{figure}[b!]
\begin{center}
\begin{tabular}{cc}
\mbox{\epsfig{figure=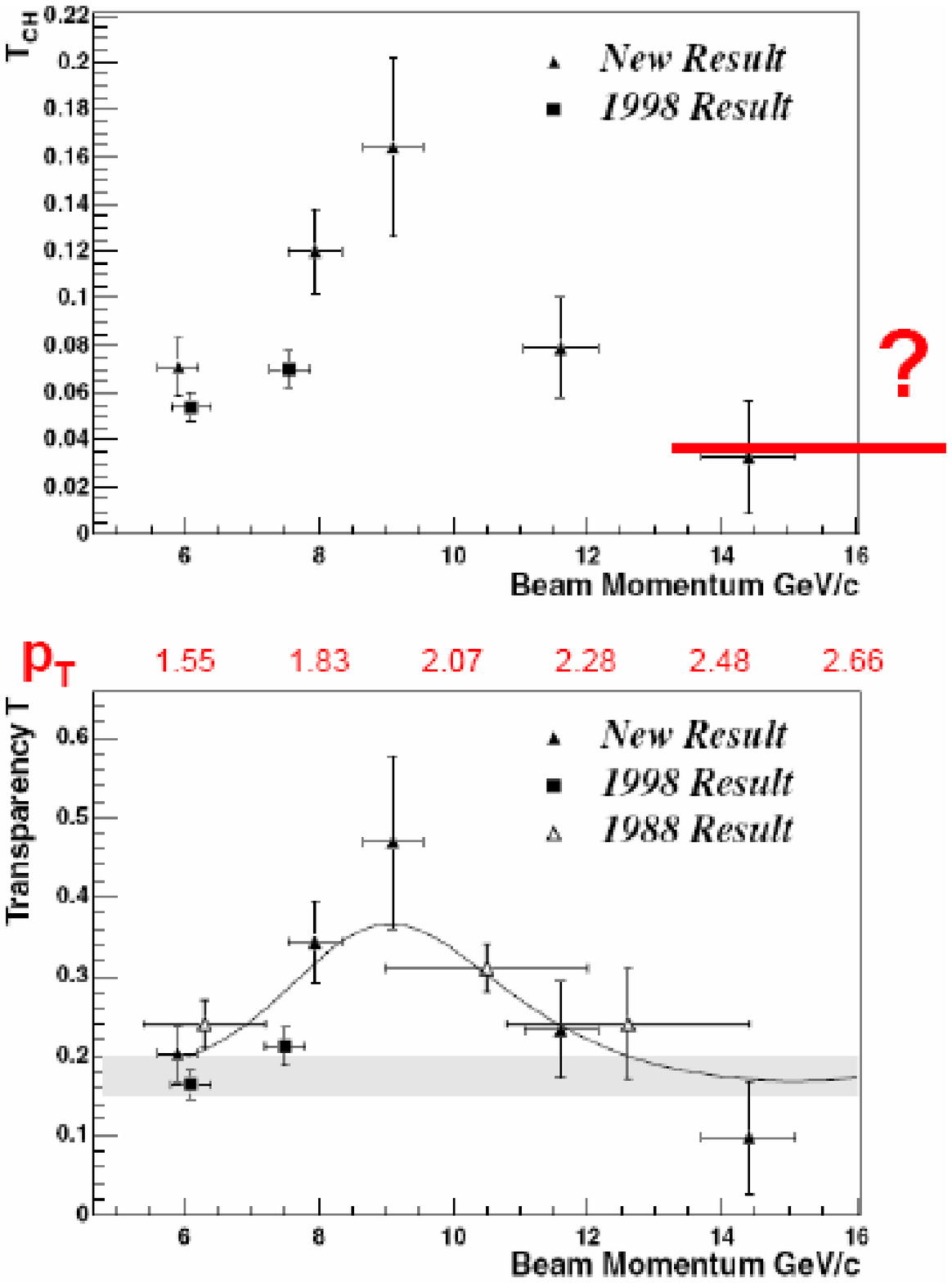,width=6cm,height=8cm}}&
\mbox{\epsfig{figure=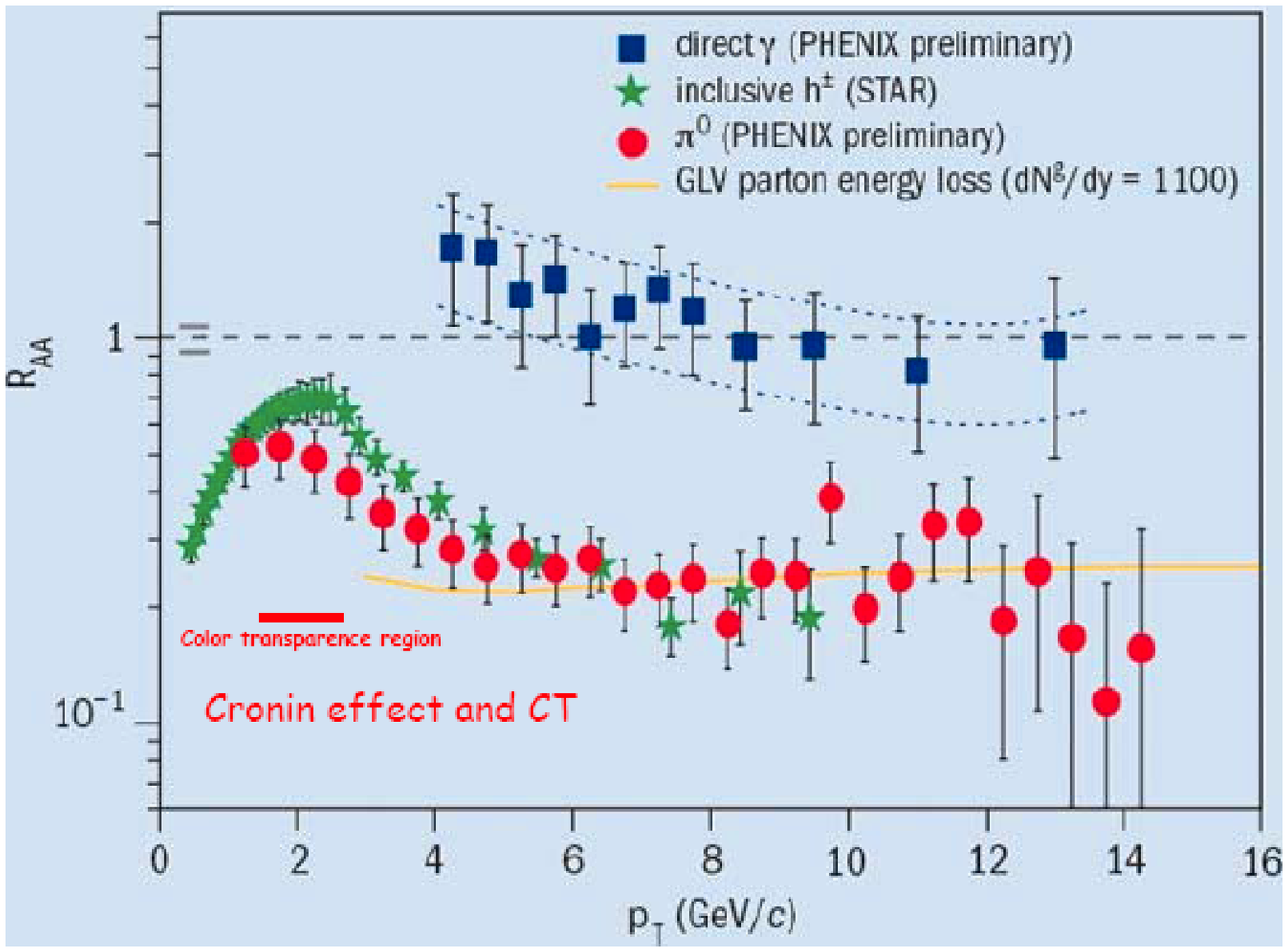,width=8cm,height=6cm}}\\
{\bf(a)}& {\bf(b)}
\end{tabular}
\end{center}
{\small{\bf Figure 1a.} The CT data from \cite{CT}.}
{\small{\bf Figure 1b.} RHIC data for "jet quenching" effect.}\\
\end{figure}

As the first step to realize this programm we will need to know
details of the spin dynamic in the nuclotron-M.

Complete description of spin dynamic in circular accelerators can
be realized using concept of a periodic precession axis
$\vec{n}(\theta)$, which is periodical function of generalized
azimuth $\theta$:  $\vec{n}(\theta)=\vec{n}(\theta+2\pi)$
\cite{bibDAN70,bibJTF71}.

Spin motion on the equilibrium orbit is a precession around the
axis~$\vec{n}$: the spin projection $J=\vec{s}\,\vec{n}$ on the
axis $\vec{n}$ will be conserved and a transversal projection
to~$\vec{n}$ is turn to the angle $\Psi=2\pi\nu$. Spin frequency
$\nu$ is shown the turn number of the particle spin during one
turn of particle in an accelerator. In traditional accelerator
with the  transverse master field (nuclotron is the accelerator
this type) the precession axis $\vec{n}$ is parallel to the
vertical axis. The spin frequency $\nu$ will changing in
proportion to the particle energy: \(\vec{n}=\vec{e}_z\), \(\nu=
G \gamma \), where $\gamma$~--- the relativistic factor,
$G=(g-2)/2$~--- anomalous part of the gyromagnetic ratio. Main
characteristic to describe the collective spin motion of particle
beam is a polarization vector
$\vec\Pi=\bigl<\vec{s}\bigr>=\bigl<J\vec{n}\bigr>$ and a
power  of depolarization $D=D=1-|\vec{\Pi}|$. The angular brackets
define that we take averaging over particle distribution in the
beam.

The motion of particles on non-equilibrium orbits give deviation
(spread) of precession axes $\Delta\vec{n}$ and spread spin
frequencies~$\Delta\nu$. If we inject beam of polarized particles
in to the nuclotron with  a spin perpendicular to the precession
axis $\vec{n}$ during the "time"\   $\theta\sim1/\Delta\nu$ (for
the nuclotron it is a some hundreds tunes) will be full
randomization spin directions related to the axis $\vec{n}$ and
polarization will be lost fully. Therefore we need to match the
polarization vector of the beam with the direction precession
axis $\vec{n}$ (vector of polarization must be parallel to the
axis $\vec{n}$). The existing channel of transportation have not
this coordination. After the ion source the polarization vector
is directed to the vertical direction. The vector polarization is
not changing direction in the linac. During transportation to the
nuclotron the rotation of polarization vector take place in
vertical and in horizontal planes. As a result the direction of
the vector of polarization will have the angle $\alpha_z$ with
vertical axis ( see Table~\ref{t:inj}).
 The power of
depolarization for not correct matchings is
$$
D_{inj}=2\sin^2\frac{\alpha_z }{2}\,.
$$

For eliminating this effect will be enough, for example, install the pair of the solenoids  at the beginning and the end of the transport
channel  which do not influence the particle trajectory and same time will turn polarization vector to the vertical line.

\begin{table}[htbp]
\begin{center}
\begin{tabular}{|@{\ \ }l|c|c|c|c|}
\hline
\hline
$\vphantom{{d_y^(d)}^(d)}$ & ${}^1H$&${}^2H$&${}^3H$&${}^3He$\\
\hline
\hline
$\vphantom{{d_y^(d)}^(d)}$ $\alpha_z$, degree& 67&9.8&116&79\\
\hline
$\vphantom{{d_y^(d)}^(d)}$ $D_{\text{inj}}$, \%& 62&1.5&55&81 \\
\hline
\hline
\end{tabular}
\parbox[t]{\textwidth}{
\caption{The power  of the beam depolarization at some mismatching
of polarization during injection in the nuclotron.}
\label{t:inj}
}
\end{center}
\end{table}

Degree of polarization in the process of acceleration can changes in region of the spin resonance, when spin frequency becomes equal to

\begin{equation}
\nu=\nu_k\,,
\qquad
\nu_k= k+k_z\,\nu_z+k_x\,\nu_x+k_\gamma\,\nu_\gamma \,.
\end{equation}
where  $\nu_x$ and $\nu_z$ are betatron frequencies, $\nu_\gamma$ is frequency of synchrotron motion. The values of betatron frequencies
are equal $\nu_x=6.8$, $\nu_z=6.85$ for the nuclotron.

The most strongest there are resonances of linear approximation, which include intrinsic resonances and resonances of structural
imperfections: integer, nonsuperperiodical and the coupling resonances of $x$ and $z$ oscillations.

Intrinsic resonances appear when spin interact with the betatron motion. Remaining resonances are connected with the distortion of the
magnetic structure of the rings which are caused by inaccuracies in production and misalignment  of the structural elements, with the
nonlinear effects of spin and orbital motions, with switching of corrective and functional elements(dipoles, quadrupoles, sextupoles and
s.o.).

Table~\ref{t:reson} shows the number of linear resonances for different particle beams ${}^1H$, ${}^2H$, ${}^3H$, ${}^3He$ in the nuclotron
($k$ and $m$ --- integer, $p=8$ --- number of superperiods).

\begin{table}[htbp]
\begin{center}
\begin{tabular}{|@{\ \ }l|l|c|c|c|c|}
\hline
\hline
Resonance type& Resonance condition&
\multicolumn{4}{|c|}{$\vphantom{{d_y^(d)}^(d)}$Number of resonances}\\
\cline{3-6}
&
$\vphantom{{d_y^(d)}^(d)}$ & ${}^1H$&${}^2H$&${}^3H$&${}^3He$\\
\hline \hline
Intrinsic resonances&
$\vphantom{{d_y^(d)}^(d)}$ $\nu=k\,p\pm\nu_z$& 6&---&8&9 \\
\hline
Integer resonances&
$\vphantom{{d_y^(d)}^(d)}$ $\nu=k$ & 25&1&32&37 \\
\hline
Nonsuperperiodical resonances&
$\vphantom{{d_y^(d)}^(d)}$ $\nu=k\pm\nu_z\,(k\neq m\,p)$& 44&2&55&64 \\
\hline
Coupling resonances  &
$\vphantom{{d_y^(d)}^(d)}$ $\nu=k\pm\nu_x$ & 49&2&63&73 \\
\hline\hline
\end{tabular}
\caption{Linear resonances in the ring of the nuclotron.} \label{t:reson}
\end{center}
\end{table}

The spin frequency grows proportionally to energy with
acceleration of beam and the intersection of spin resonances
becomes unavoidable. The basic parameters for crossing the spin
resonance are the  spin resonance strength $w_k$, detuning from
the resonance $\varepsilon=\nu-\nu_k$ and speed of detuning
changing $\varepsilon'={d\varepsilon}/{d\theta}$ (speed of
crossing). The spin resonance strength $w_k$ is the
corresponding Fourier-harmonic of transverse spin disturbance
$\vec{w}$ and determines the width of dangerous interval in
region of the spin resonance.

We can distinguish three possibility to cross the resonance with constant speed there are fast, adiabatic and intermediate crossings. The
beam practically completely will be depolarized with the intermediate crossing of resonance ( $|w_k|^2\sim \varepsilon'$).With the fast
intersection ($|w_k|^2 \ll \varepsilon'$) the polarization vector $\vec{\Pi}$  hasn't time to considerably change and the degree of
depolarization is equal to \( D\simeq(\pi\,\left< |w_k|^2\right>)/{\varepsilon }' \). With the slow (adiabatic) crossing when ($|w_k|^2 \gg
\varepsilon'$) take place overturn of the polarization vector relative to the vertical direction. In this case should be distinguished the case of
"coherent" \ and "incoherent" \ crossing. "Coherent" \ crossing means that the resonance strength is identical for all particles (integer
resonances). In this case the condition ($w^2_k \gg \varepsilon'$) is satisfied for all particles of the beam and degree of polarization
after crossing remains with exponential accuracy. With the "incoherent" \ crossing the resonance strength is different for different
particles and, for example, it depends on the amplitude of betatron oscillations (intrinsic resonances). There are not only particles with adiabatic
type crossing in the beam, but the intermediate and fast types of crossing, which leads to the partial depolarization of the
beam. With the normal distribution of particle coordinates and momentums in the beam the degree of depolarization will be equal:
\(D\simeq{\varepsilon }'/(\pi\,\left< |w_k|^2\right>)\). With the adiabatic crossing it is necessary to consider the synchrotron
oscillations of the particles, whose accounting can lead to the partial or even complete depolarization.

It is  convenient for calculations to introduce new  parameter as the characteristic
 resonance strength \(w_d=\sqrt{{\varepsilon }'/\pi}\).  Intersection of the spin resonance lead to
practically the complete depolarization of the beam when the resonance strength is equal to $w_d$. Then the
resonance strength, which corresponds to loss by 1\% of polarization with
the fast crossing, is equal to $0.1\,w_d$, and the resonance strength, which corresponds to loss by 1\% of polarization with
the adiabatic crossing, is equal to $10\, w_d$ ("incoherent"\
resonances), $3.26\, w_d$ ("coherent"\ resonances).

The results of calculation of main characteristics for crossing of the spin
resonances and their strengthes are given in  Table~\ref{t:resparam} \cite{bibDeuteronNucl,bibProtonNucl}.

\begin{table}[htbp]
\begin{center}
\begin{tabular}{|@{\ \ }l|c|c|c|c|}
\hline
\hline
$\vphantom{{d_y^(d)}^(d)}$ & ${}^1H$&${}^2H$&${}^3H$&${}^3He$\\
\hline
\hline
$\vphantom{{d_y^(d)}^(d)}$ G&1.793&-0.143& 7.92&-4.184\\
\hline
$\vphantom{{d_y^(d)}^(d)}$ $E_k^{\max} \,,\  \text{[GeV/u]}$&12.84&6.00&3.74&8.28 \\
\hline
$\vphantom{{d_y^(d)}^(d)}$ $\nu_{min}$ -- $\nu_{max}$ & 1.8 -- 26.3 &-1.05 -- -0.144&7.92 -- 39.5&-41.1 -- -4.19\\
\hline
$\vphantom{{d_y^(d)}^(d)}$ $\varepsilon^\prime$, $\left(\tau_{\text{accel}}=0.5 s\right)$&
$7.0\cdot 10^{-6}$&$2.8\cdot 10^{-7}$&$1.0\cdot 10^{-5}$&$1.1\cdot 10^{-5}$\\
\hline
$\vphantom{{d_y^(d)}^(d)}$ $w_d$, $\left(\tau_{\text{accel}}=0.5 s\right)$&
$1.5\cdot 10^{-3}$&$3.0\cdot 10^{-4}$&$1.8\cdot 10^{-3}$&$1.9\cdot 10^{-3}$\\
\hline
\hline
\end{tabular}
\caption{Crossing characteristics  of the spin resonances in the
nuclotron.}
\label{t:resparam}
\end{center}
\end{table}

\begin{figure} [p]
\centering
\vskip -10mm
\includegraphics[width=0.49\textwidth]{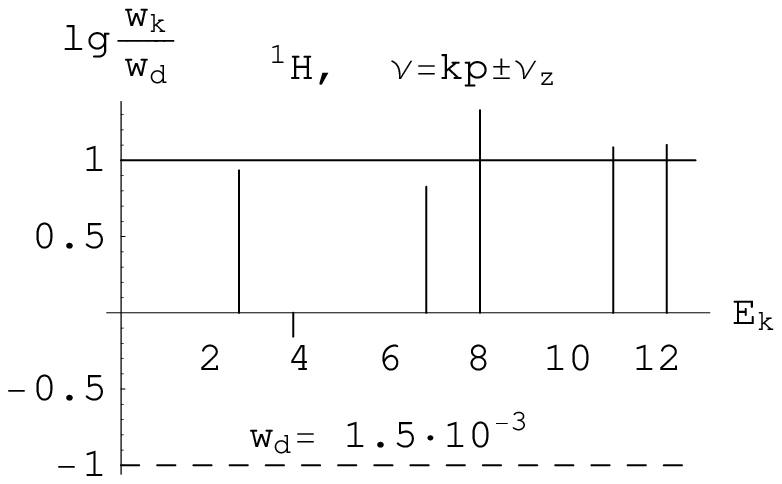} \hfill
\includegraphics[width=0.49\textwidth]{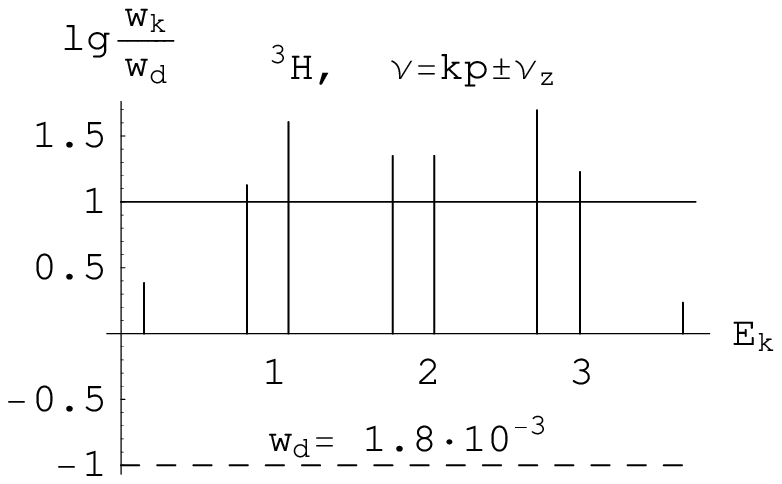}\\
\vskip 2mm
\includegraphics[width=0.49\textwidth]{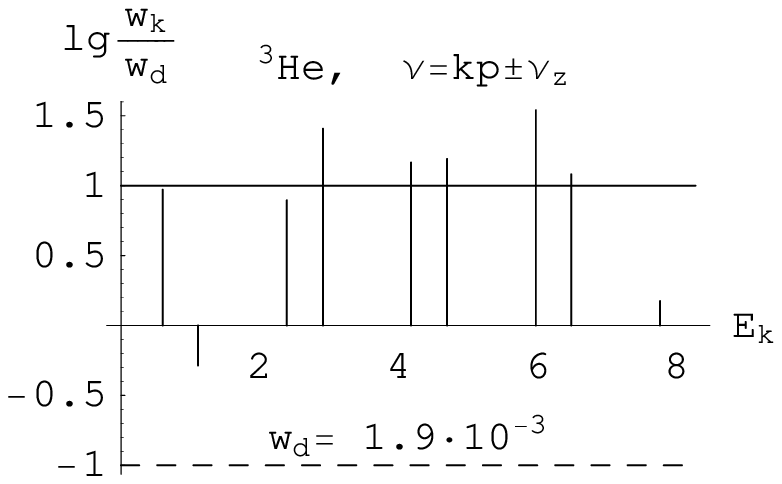}
\\
\vskip -3mm
\parbox[t]{0.9\textwidth}{
 \caption{Intrinsic resonances.} \label{f:intrisic}
}
\vskip 5mm
\includegraphics[width=0.49\textwidth]{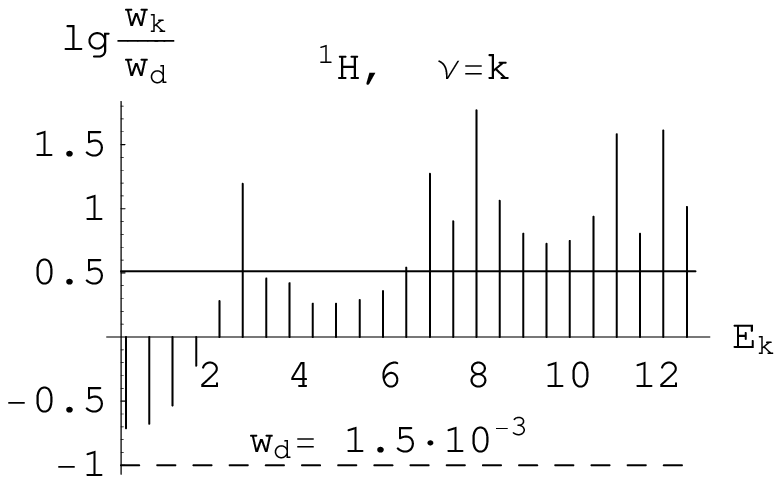} \hfill
\includegraphics[width=0.49\textwidth]{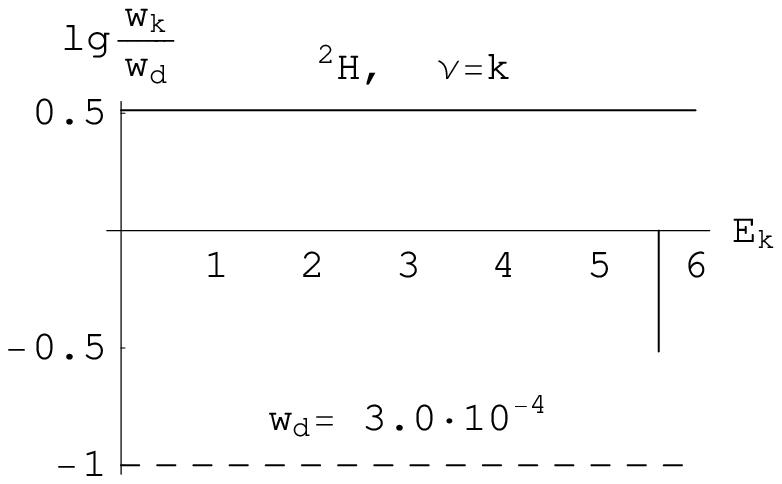}\\
\vskip 2mm
\includegraphics[width=0.49\textwidth]{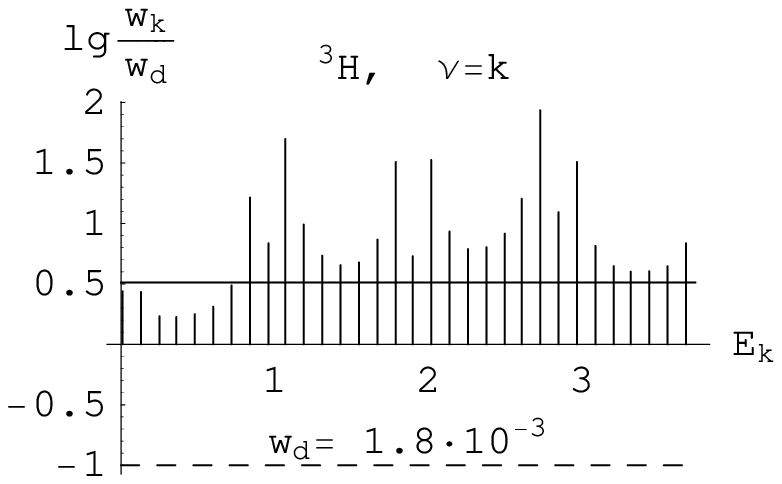} \hfill
\includegraphics[width=0.49\textwidth]{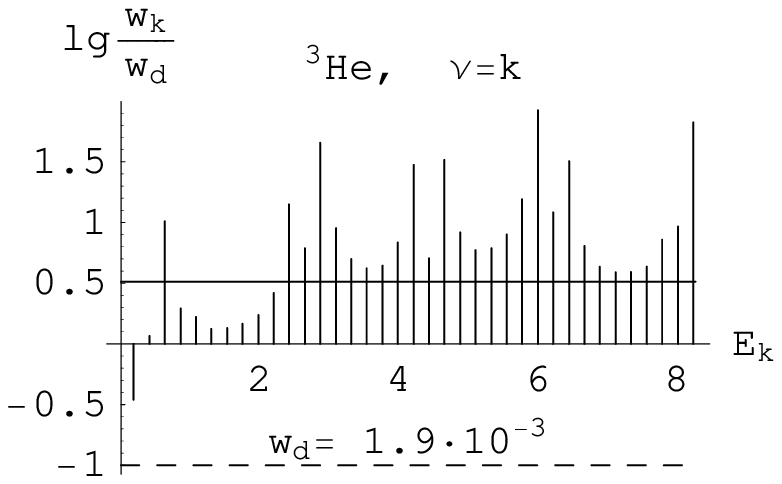}\\
\vskip -3mm
\parbox[t]{0.9\textwidth}{
 \caption{Integer resonances.} \label{f:int}
}
\end{figure}

Figures~\ref{f:intrisic}-\ref{f:int} show the
logarithmic graphs of the resonance strengthes of linear
approximation in units of the characteristic strength $w_d$ in the
operating range of beam kinetic energy $E_k$. Each graph is split
into three regions, which correspond to the intermediate crossing
(region between the continuous and dotted lines), fast crossing
(under the dotted line) and adiabatic crossing (above the solid
line). It was assumed in calculation of the resonance strengthes that
the emittances in the horizontal and vertical direction at the
energy of injection there are equal $45\pi\,\mbox{mm} \cdot \mbox{mrad}$,
adjustment errors of quadrupoles --- $0.1$~mm and adjustment
errors of the turning for main magnets --- $0.001$~rad.

The resonances located in the zone of intermediate crossing lead
to the depolarization of the beam. From the comparison of graphs
it follows that almost in full of energy range the
depolarization take place for intrinsic and integer resonances
\mbox{(Figure~\ref{f:intrisic}-\ref{f:int})}. The coupling resonances
and nonsuperperiodical resonances also can lead to
the depolarization of the beam in the same regions of energy
where intrinsic resonances are located.

Let us consider methods of crossing of the spin resonances the
most suitable for the nuclotron. In crossing of the integer
resonances with the intermediate strength ($|w_k|^2\sim
\varepsilon'$) it is expedient to use a method of premeditated
increasing of  the resonance strength~\cite{bibJTF71}. For this
purpose it is enough to insert in free nuclotron gaps some
longitudinal magnetic field. The resonance strength with this
longitudinal field is determined by expression
$$
w_k=\frac{\varphi_y}{2\pi}=\frac{(1+G) H_y L_y}{2\pi HR}
$$
and must be correspond to the condition of the adiabatic crossing
\( |w_k|^2\gg \varepsilon'\). Furthermore in order to avoid the
effects of depolarization because of synchrotron modulation of
energy necessary to satisfy also the condition: $ |w_k|^2\gg
\sigma \,\nu_\gamma\sim  10^{-2}$ , where
$\sigma=\nu\sqrt{\bigl<(\Delta \gamma/\gamma)2\bigr>}$ --- the
amplitude of synchrotron modulation of energy, and $\nu_\gamma$
--- the frequency of synchrotron oscillations \cite{bibProtonNucl}.

The maximal values of
integrals of the longitudinal field (on the energy of extraction)
which need to guarantee the adiabatic crossing of the integer
resonances in full range of the energy are given in Table~\ref{t:adiabcross}.

\begin{table}[htbp]
\begin{center}
\begin{tabular}{|@{\ \ }l|c|c|c|c|}
\hline
\hline
\vs & ${}^1H$&${}^2H$&${}^3H$&${}^3He$\\
\hline
\hline
\vs $(H_y L_y)$, \ $\text{T}\cdot\text{m}$ &$1$ &$3.4$&$0.3$&$0.9$\\
\hline\hline
\end{tabular}
\caption{The integrals of the longitudinal field for the adiabatic
crossing.}
\label{t:adiabcross}
\end{center}
\end{table}

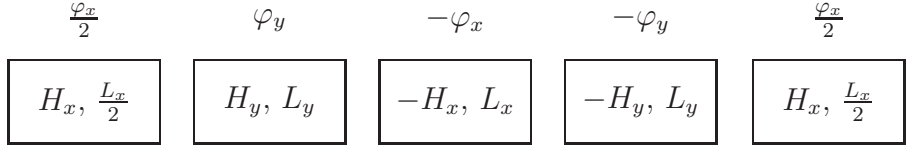
\begin{figure}[htb]\centering
\unitlength=1.00mm
\special{em:linewidth 0.4pt}
\linethickness{0.4pt}
\begin{picture}(126.00,25.00)(0,13)
\put(5.33,18.67){\framebox(20.00,11.00)[cc]{$H_x$, $\frac{L_x}2$}}
\put(15.33,35.00){\makebox(0,0)[cc]{$\frac{\varphi_x}2$}}
\put(30.00,18.67){\framebox(20.00,11.00)[cc]{$H_y$, $L_y$}}
\put(40.00,35.00){\makebox(0,0)[cc]{$\varphi_y$}}
\put(54.67,18.67){\framebox(20.00,11.00)[cc]{$-H_x$, $L_x$}}
\put(64.67,35.00){\makebox(0,0)[cc]{$-\varphi_x$}}
\put(79.33,18.67){\framebox(20.00,11.00)[cc]{$-H_y$, $L_y$}}
\put(89.33,35.00){\makebox(0,0)[cc]{$-\varphi_y$}}
\put(104.33,18.67){\framebox(20.00,11.00)[cc]{$H_x$, $\frac{L_x}2$}}
\put(114.33,35.00){\makebox(0,0)[cc]{$\frac{\varphi_x}2$}}
\end{picture}
\vskip -5mm
\caption{The "insert" to control  the spin frequency.}
\label{f:strjump}
\end{figure}

When crossing the resonances with the betatron frequencies it is
possible to use a method of compensation the degree of
depolarization \cite{bibCrossSpinRes1}. Conservation of the
degree of polarization is ensured due to control of detuning
$\varepsilon=\nu-\nu_k$ inside the  resonance region. The control
of detuning  $\varepsilon$ during crossing is possible due to
changing the spin frequency $\nu$. For this it is necessary to
introduce into the ring of nuclotron an "insert" \ with an
additional magnetic field  which makes it possible to obtain the
required dependence of the spin frequency on a magnetic field
$\nu=\nu(\vec{H})$. There is possible to use the "insert" with
the longitudinal and radial fields, depicted in
Figure~\ref{f:strjump}, where \mbox{$\varphi_x$, $\varphi_y$ ---
} angles of spin turns around the radial and the longitudinal
fields.

In the approximation of a small spin angular turn
($\varphi_x,\varphi_y\ll 1$) the direction of the equilibrium
polarization remains vertical and changing of the precession
frequency of the spin is equal to $ \Delta \nu =
(\varphi_x\varphi_y)/({2\pi})$. The maximal vertical deviation of
the equilibrium orbit caused by radial fields will be \( \Delta
z_{\max}=\varphi_x/({8\nu})(4L_y+5L_x)\, \). The maximal length
of the "insert" \ is limited by the length of free space in the
accelerator which in the nuclotron is about  $350$~cm.

The depolarization of beam is possible during the slow beam
extraction from the nuclotron when the energy of the beam close
to energy of the spin resonances. The degree of depolarization in
this case depends on the spin resonance strength $w_k$ and
the detuning from the resonance $\varepsilon$. For the completely
polarized beam at the beginning power of depolarization will be: \( D \simeq
\dfrac{\bigl<w^2_k\bigr>}{2\,\varepsilon^2} \). In this case to
avoid depolarization one need move away from the resonance to the
value $\Delta\varepsilon\sim 10 w_k \,(\Delta \gamma
=\Delta\varepsilon/G)$. For example, for the beam of protons this value will
be $\Delta \gamma\simeq 50\text{MeV}$ for the detuning from the
resonance with the strength $w=10^{-2}$ (adiabatic crossing) and
$\Delta \gamma\simeq 5\text{MeV}$ for the dutuning from the
resonance with the strength  $w=10^{-3}$ (intermediate crossig).

\end{document}